# Investigation on Gait Time Series by Means of Factorial Moments

Huijie Yang[1,+]，Fangcui Zhao[1]，Yizhong Zhuo[2]，Xizhen Wu[2]

[1] Physics Institute, Hebei University of Technology, Tianjin 300130, China

[2] China Institute of Atomic Energy, Beijing 102413, P.O.X 275(18), China

**Abstract**

By means of Factorial Moments (FM), the long-range correlations embedded in gait time series are investigated. It is found that FM is an effective tool to deal with this kind of time series.

In recent years it has been recognized that in many nature sequences the elements are not positioned randomly, but exhibit long-range correlations. Typical examples include non-coding DNA sequences [1,2], weather records [3], as well as sequences of heartbeat/gait rate [2,4,5,6] and information stream in networks. The common feature that the long-range correlations decay by a power law may be useful, e.g., in DNA for distinguishing between coding and non-coding sequences, in atmosphere science for testing state-of-art climate models and in clinicians for diagnosing disease.

In literatures detailed discussions on gait time series have been presented using the method of DFA [6]. In this paper, for the first time we investigate gait time series by means of the concept of factorial moments.

## I. Gait Time Series and DFA Method

For a walker we can measure the time interval between two successive strides with a high precision. And a large amount of the time intervals, namely gait intervals, form a time series. The fluctuations of gait interval values around the average over a period constitute a body of data that attract special interests to statistical physicists. When we begin to study these admirably accurate data, almost immediately we encounter three roadblocks [5,7]. The first is that we must find the non-trivial self-similar properties instead of the trivial self-similar ones. Scaling the t (time) dimension only and keeping the y (time interval) dimension invariance, we can obtain segments of gait interval series at different time scales. The segments have self-similar property. But it is a trivial one that we are not interested in. The second is that during the measurements a walker (an object) is perturbed by environment always. And the perturbations can be described with white noise. What is more, statistical noise appears due to the finite number of measurements. The third one is that these data have the property of non-stationary. This means that the statistical properties of gait time intervals are not constant in time; they are not served up neatly as independent, identically distributed random variables. And non-stationary is the essential problem

---

[+] Corresponding author　　E-mail: huijieyangn@eyou.com

to be resolved.

Normally, to find the non-trivial self-similar properties and reduce the perturbations due to noise, an integrated time series is introduced instead of investigating the gait series directly. The integrated time series can be constructed as follows,

$$Y_m = \sum_{i=1}^{m} \Delta\tau_i$$

Where $\{Y_m | m = 1,2,3,...\}$ forms the integrated time series, called $Y_m$ "profile", and $\{\Delta\tau_i | i = 1,2,3,...\}$ is the initial gait interval series.

To avoid spurious detection of correlations due to artifact of non-stationary, DFA method is employed to calculate the time-dependent fluctuation function $F(n)$ up to fourth order. The profile $Y_m$ is divided into non-overlapping segments of size *n*. In first order DFA (DFA1) the best linear fit of the profile in each segment is determined, and the standard deviation of the total profile from these straight lines represents the fluctuation function $F(n)$. Linear trends in the profile are eliminated by this procedure. In second order DFA (DFA2) the best quadratic fit of the profile in each segment is determined, and the standard deviation of the total profile from these parabolas represents the fluctuation function $F(n)$. The high order DFAs are straight-forward extensions: In *q*th order DFA, trends of order *q* in the total profile and of order *q-1* in the original data are eliminated.

The DFA method is designed by C. -K. Peng et al., and systematical works have been done with this method itself and its applications to biology, medicine and economics, etc. [5-14].

## II. Factorial Moments (FM)

More than ten years have witnessed a remarkably intense experimental and theoretical activity in search of scale invariance and fractal in multihardron production processes, for short also called "intermittency" [15]. The primary motivation is the expectation that scale invariance or self-similarity, analogous to that often encountered in complex non-linear systems, might open new avenues ultimately leading to deeper insight into long-distance properties of QCD and the unsolved problem of colour confinement.

Generally, intermittency can be described with the concept of probability moment (PM). Dividing a region of phase space $\Delta$ into *M* bins, the volume of one bin is then $\delta = 1/M$. And the definition of *q*-order PM can be written as [16],

$$C_q(\delta) = \sum_{m=1}^{M} p_m^q,$$

Where $p_m$ is the probability for a particle occurring in the $m$'th bin, which satisfies a constrained condition, $\sum_{m=1}^{M} p_m = 1$. For a self-similar structure, PM will obey a power law as,

$$\lim_{\delta \to 0} C_q(\delta) \propto \delta^{(q-1)D_q},$$

And $D_q$ is called $q$-order fractal dimension or Renyi dimension. Simple discussions show that $D_0, D_1$ and $\{D_q | q \geq 2\}$ reflect the geometry, information entropy and particle correlation dimensions, respectively.

It is well known that intermittency is related with strong dynamical fluctuations. But the measurements for multihardron production obtain the distribution of particle numbers directly instead of the probability distribution. And the finite number of cases will induce statistical fluctuations. To describe the strong dynamical fluctuations and dismiss the statistical fluctuations effectively, factorial moment (FM) is suggested to investigate intermittency [17,18]. The generally used form for FM can be written as,

$$F_q = M^{q-1} \sum_{m=1}^{M} \frac{<n_m(n_m-1)...(n_m-q+1)>}{<n>^q},$$

Where M is the number of the bins the considered interval being divided into, $n_m$ the number of particles occurring in the $m$'th bin, and $n$ the total number of particles in all the bins. A measure quantity can then be introduced to indicate the dynamical fluctuations,

$$\phi_q = \lim_{\delta \to 0} \frac{\ln F_q}{\ln(1/\delta)}.$$

For a self-similar structure, $\phi_q$ will be a non-zero value, and non-linear strong dynamical fluctuations exist.

Here we present a simple argument for the ability of FM to dismiss statistical fluctuations due to finite number of cases [16].

The statistical fluctuations will obey Bernoulli and Poisson distributions for a system containing uncertain and certain number of total particles, respectively. For a system containing uncertain total particles, the distribution of particles in the bins can be expressed as,

$$Q(n_1, n_2, ... n_M | p_1, p_2, ... p_M) = \frac{n!}{n_1! n_2! ... n_M!} p_1^{n_1} p_2^{n_2} ... p_M^{n_M}.$$

And $(p_1, p_2, ...p_M)$ are the probabilities for a particle occurring in the $1, 2, ..., M$ bins, respectively. Hence,

$$\langle n_m(n_m-1)...(n_m-q+1) \rangle$$
$$= \int dp_1 dp_2...dp_M P(p_1, p_2,...p_M) \times \sum_{n_1}...\sum_{n_M} Q(n_1, n_2,...,n_M | p_1, p_2,...p_M)$$

$$\times n_m(n_m-1)...(n_m-q+1)$$

$$= n(n-1)...(n-q+1) \times \int dp_1 dp_2...dp_M P(p_1, p_2,...,p_M) p_m^q$$

$$= n(n-1)...(n-q+1) \langle p_m^q \rangle.$$

That is to say,

$$F_q(M) = C_q(M) \propto M^\phi, |M \to \infty.$$

Therefore FM can describe the strong dynamical fluctuations and can dismiss the statistical fluctuations effectively.

Besides the statistical fluctuations, there are some trivial dynamical processes that need to be dismissed. These trivial dynamical processes induce the average numbers of particles in different phase space bins being not same, and the form of FM should be the original one, which reads,

$$F_q = M^{-1} \sum_{m=1}^{M} \frac{<n_m(n_m-1)...(n_m-q+1)>}{<n_m>^q},$$

A typical method to dismiss the fluctuations due to this kind of trivial dynamical processes is to transform the original distribution to homogeneous distribution by means of integrate method as follows [19],

$$x(y) = \frac{\int_{y_a}^{y} f(y)dy}{\int_{y_a}^{y_b} f(y)dy}.$$

But in this paper we resolve this problem by constructing a series of delay register vectors based upon the gait time series.

To find self-similar structures embedded in a gait time series by means of FM, a process can be constructed as [20-24],

(1). d successive gait intervals along the gait series we are interested in are regarded as the state of a case containing d particles. The state of the case can be described with a *d*-dimensional vector as $(x_1, x_2, x_3...x_d)$, where $x_i$ is the state value for the $i'th$ gait interval.

(2). The total possible $N-d+1$ successive cases form a process. The process covers the entire gait time series we are interested in, which can be expressed with a series in d-dimensional **delay-register vectors:**

$$(x_1, x_2, x_3 \ldots x_d)$$

$$(x_2, x_3, x_4 \ldots x_{d+1})$$

$$\downarrow$$

$$(x_{N-d+1}, x_{N-d+2}, x_{N-d+3} \ldots x_N)$$

As a summary, we can illustrate several features about FM as follows,
(1). By means of constructing delay-register series we can dismiss trivial dynamical processes we are not interested in at all. And similar with the integrate transformation in DFA method, we can find nontrivial self-similar properties and eliminate statistical fluctuations due to finite number of records in a certain degree.
(2). FM can dismiss the statistical noises due to finite number of records and environments. FM can extract and describe strong dynamical fluctuations that induce self-similar structures almost exactly.
(3). The most important feature for FM is that it can resolve the non-stationary problem in a certain degree. FM is determined by the distribution of particle numbers in all the phase space bins. And same numbers of particles in different bins will bring same contributions to FM, though they can induce different values of ordinary statistical quantities. FM can overcome the non-stationary problem.

Hence, we can expect that FM may be a powerful tool to describe the long-range correlations in gait time series.

## III. Applications to Gait Time Series Analysis

The concept of FM has been used to deal with many kinds of complex dynamical processes in physics, such as multi-particle production at high energy, DNA melting and denaturalization with the temperature increasing, etc. [25-26]. What is more, this concept is also improved to a new version called etermittency, to deal with some problems where statistical average can not be complemented properly [27].

At the beginning, FM is used to deal with many kinds of dynamical processes, and fruitful results are obtained. But FM is mainly restricted to one-dimensional variables, and FM tends to saturate at small experimentally allowed resolution, instead of power-law. It is believed that power-law exists in high dimensional phase space. Measurements for high dimensional phase space show that for two-dimensional phase space, FM varies a lot with the plane of projection, while for three-dimensional phase space FM is a smoothly upward bending curve. Then FM is analyzed with self-affine instead of self-similar for the shrinkage ratios for three directions being different and upward bending is dismissed effectively.

For a gait time series there are two dimensions with different scales, time and time interval, respectively. Therefore it is different with a curve in a 2-dimensional plane completely. Here we perform scale transformations for the two directions simultaneously. That is to say, considering a phase space region that can be written as $(t_1 \to t_2, \Delta t_{min} \to \Delta t_{max})$, we divide the interval $\Delta t_{min} \to \Delta t_{max}$ into 2,4,8,16,32, 64, 128 bins, respectively. And consider different windows along t-axis to construct cases, such as 4,8,16,32,64,128,256,and 512, respectively.

As an example we consider three objects, e.g., a healthy elder (77 year-old), a healthy young (23-year-old) and an object suffering Parkinsonism [28]. Fig.(1) shows the results for the healthy elder object. And only the direction of $\Delta t$ is re-scaled. FM values versus different sizes of windows are presented. When the sizes of windows are small enough, we can obtain a typical relation, $LnF_q \propto LnM$. For other sizes of windows, FM tends to saturate. In Fig.(2) the results for a Parkinsonism object is presented. We can find that FM tends to saturate for different sizes at all. Hence performing scale transformation for two directions simultaneously is an essential step to find fractal structures for gait time series, just as presented in Fig.(3). Here the window sizes and the number of bins $M$ are converted into the number of phase space boxes in two dimensions according to a general relation, e.g., $N_{box} = C \times \frac{M}{S_{windows}}$. C is a constant, and is set to be 1000 in this paper. M and $S_{windows}$ are the numbers of bins the interval $\Delta t_{min} \to \Delta t_{max}$ being divided into and the sizes of windows along $t$ dimension, respectively. It is interesting to find that for health objects, no matter young or elder, FM obey an exact power law, e.g., $F_q \propto N_{box}^{\phi_q}$, in a wide range of scales. While for the object suffering Parkinsonism we can not find the self-similar property at all.

In summary, FM can extract the non-linear dynamics from gait time series effectively. It may be also a powerful tool to deal with many kinds of temporal series, such as heartbeat, gait and information stream in Internet, etc.

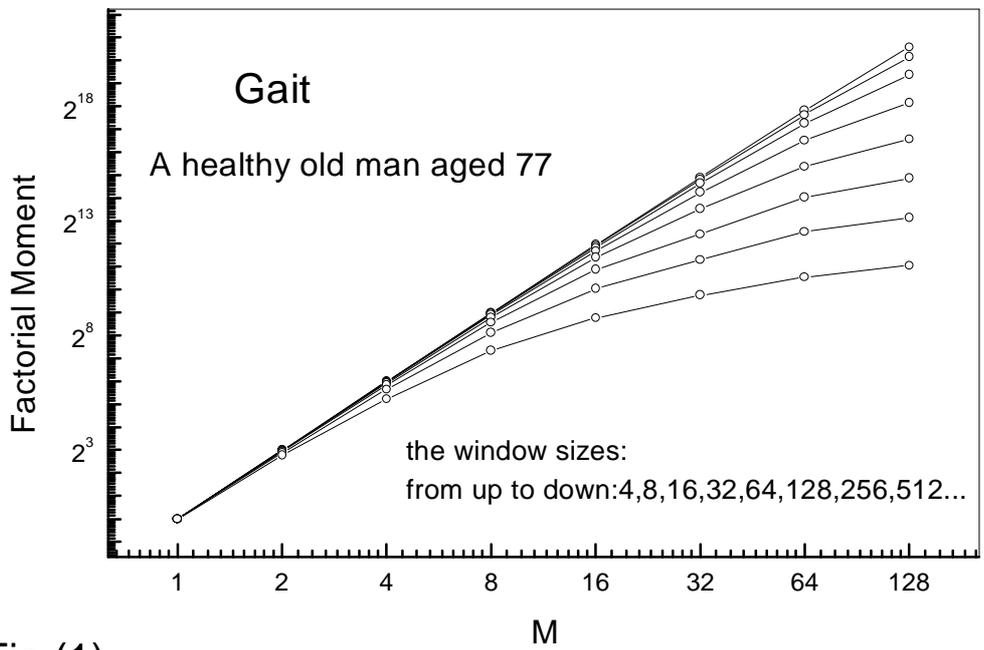

Fig.(1)　The relation between saturation and window size

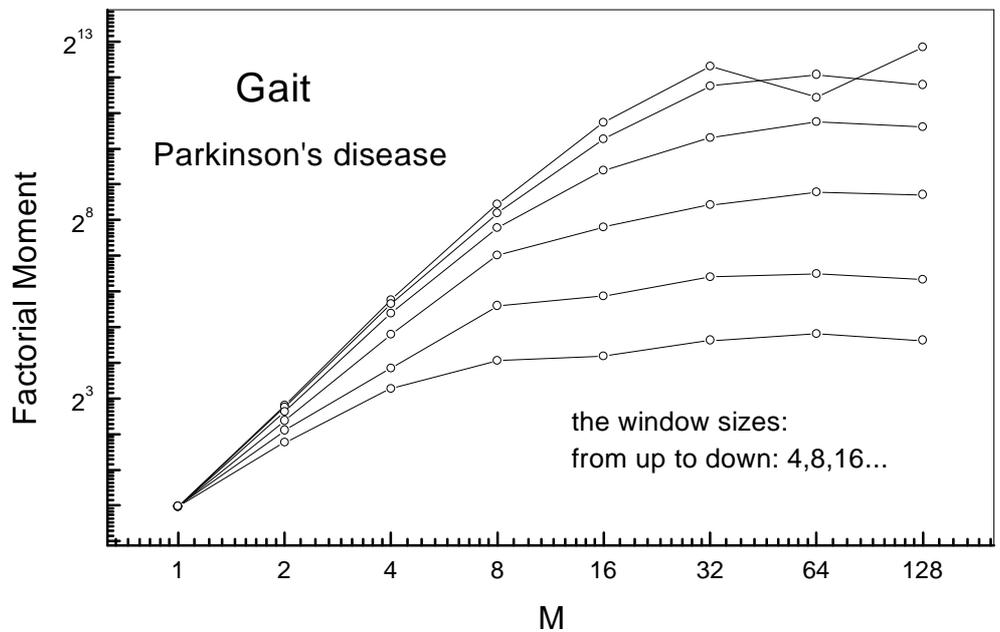

Fig.(2)　The relation between saturation and window size

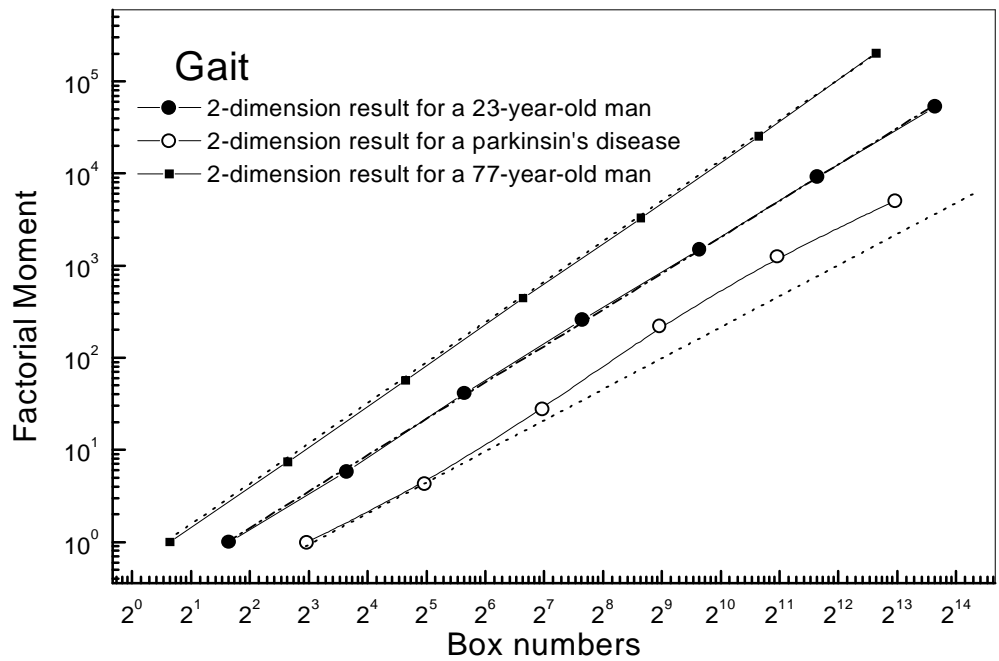

Fig.(3)    The relation between the number of boxs and FM